\documentclass[11pt]{article}
\usepackage{amsmath}
\usepackage{latexsym}
\usepackage{amssymb}
\usepackage{graphicx}

\begin{document}

\title{A Constructive Critique of the Three Standard 
Systems\footnote{Concluding talk
``LHC in Context" at the Praha-ASI Conference on LHC Physics, Prague, 
July 2003.}}
\author{F. Wilczek\footnote{Email: wilczek@mit.edu}\\
\\
{\small\itshape Department of Physics}
  \\
{\small\itshape Center for Theoretical Physics}\\
{\small\itshape Laboratory of Nuclear Science}\\
{\small\itshape Massachusetts Institute of
Technology} \\
{\small\itshape  Cambridge, Massachusetts 02139}}

\date{\small MIT-CTP-3466}

\maketitle

\pagestyle{myheadings} \markboth{F. Wilczek}{Three Standard System}
\thispagestyle{empty}

It has become conventional to say that our knowledge of fundamental
physical law is
summarized in a Standard Model.  But this convention lumps together two quite
different conceptual structures, and leaves out another.  I think it
is more accurate
and informative to say that our current, working description of
fundamental physics
is based on three standard conceptual systems.  These systems are
very different; so
different, that it is not inappropriate to call them the Good, the
Bad, and the Ugly.
They concern, respectively, the coupling of vector gauge particles,
gravitons, and
Higgs particles.  It is quite a remarkable fact, in itself, that
every nonlinear
interaction we need to summarize our present knowledge of the basic (i.e.,
irreducible) laws of physics involves one or another of these particles.

\section{The Gauge Sector}

The unambiguously good system is one describing couplings of the
$SU(3)\times SU(2)\times U(1)$ gauge
bosons.  Deep principles of symmetry and locality greatly constrain
the form of these
couplings.  When we combine these principles with the demand of
renormalizability,
we come down to a theory containing just one continuous parameter for
each gauge
group, namely its overall coupling strength.   (Strictly speaking
this is only true for
the nonabelian factors, and only if we put aside the $\theta$ terms
for those factors.
I'll return to the first of these points below; for more on the other, see \cite{spacePart}.)  This system gives us an
extraordinarily economical
account of the central features of the strong, weak and
electromagnetic interactions,
which is in excellent agreement with a host of accurate experiments.

This is not the place for yet another retelling of that story,
wonderful though it is.
Let me just invoke it with a two familiar icons, which I'll want to
refer to later.  We
should not let familiarity blind us to their beauty and power.

Figure 1 shows the running of the coupling in QCD.  It summarizes the
results of
many hundreds of independent measurements in different situations at different
energy-scales, all of which conform to the predictions of an
extremely tight theory.
Two features are especially noteworthy.  First, there is a special
point labeled ``lattice
gauge theory".  Whereas the other points are grounded in perturbative QCD
(including, to be sure, use of the renormalization group, non-perturbative
factorization theorems, and multiloop calculations), this one employs the basic
algorithmic definition, with no compromises, of the only relativistic
quantum field
theory that can withstand such use.  Second, the theoretical band of allowed
couplings focuses to the right, as we approach large energy scales.
This means that
by the time we reach the LEP experiments, our predictions are essentially
parameter-free!  Indeed, any reasonable value for the QCD $\Lambda$ 
parameter produces
the same value of $\alpha_s(M_{\rm W})$, within a few percent (and the relevant
quark masses have become, at these high energies, negligible).
\begin{figure}
\begin{center}
\includegraphics[scale=.80]{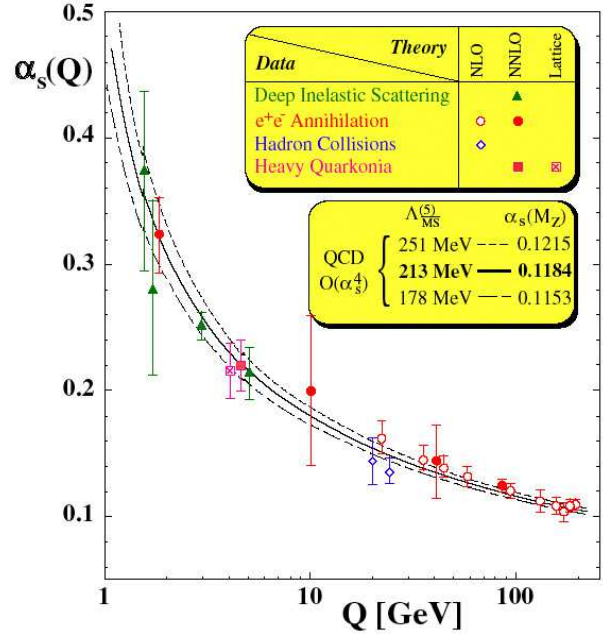}
\caption{Running of the coupling in QCD}
\end{center}
\label{fig1}
\end{figure}

Figure 2 shows the over-constrained fit of electroweak parameters to
precision measurements.  Such measurements were used to predict the mass of the
W and Z bosons and the top quark before their discovery.  Looking ahead, they
suggest that the Higgs particle will be reasonably light.  We expect
150 GeV $\leq
m_{\rm H}$ - eminently accessible at the LHC - unless of course the
particles and
interactions we know are somehow part of a larger conspiracy, which mimics the
standard model accurately in all other respects.
\begin{figure}
\begin{center}
\includegraphics[scale=.80]{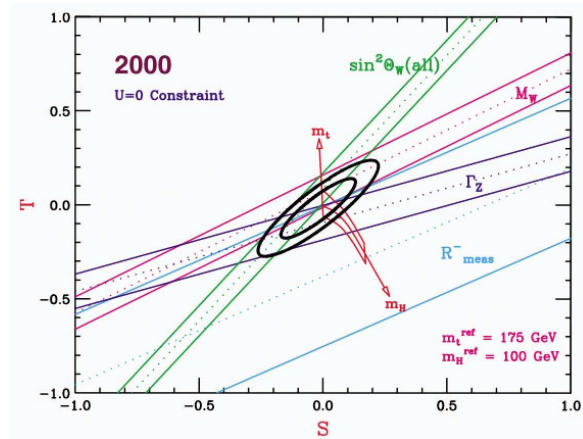}
\caption{Precision tests of electroweak theory}
\end{center}
\label{fig2}
\end{figure}

\subsection{Self-Transcendence}

The empirical success of our gauge theories, and the tight, elegant
mathematical
structure they share, is so clear and impressive that we can build upon it with
confidence.  We must take very seriously, and should
strive to remove,
the remaining esthetic flaws of these theories.

Looking critically at the structure of a single standard model
family, as displayed
in Figure 3, one has no trouble picking out flaws.

The gauge symmetry contains three separate pieces, and the fermion
representation
contains five separate pieces. While this is an amazingly tight
structure, considering the
wealth of phenomena described, it clearly fails to achieve the
ultimate in simplicity and
irreducibility.  Let me remind you, in this context, that
electroweak ``unification" is
something of a misnomer.  There are still two separate symmetries,
and two separate
coupling constants, in the electroweak sector of the standard model.
It is much more
accurate to speak of electroweak ``mixing''.

Worst of all, the abelian $U(1)$ symmetry is powerless to quantize
its corresponding charges. The hypercharge assignments -- indicated
in Figure 3 by the
numerical subscripts -- must be chosen on purely phenomenological
grounds. On the
face of it, they appear in a rather peculiar pattern. If we are
counting continuous
parameters, the freedom to choose their values takes us from three to
seven (and
more, if we restore the families).  The electrical neutrality of
atoms is a striking and
fundamental fact, which has been checked to extraordinary precision,
and which is
central to our understanding of Nature.  In the standard model this
fact appears, at a
classical level, to require finely tuned hand-adjustment.
\begin{figure}
\begin{center}
\includegraphics[scale=.40]{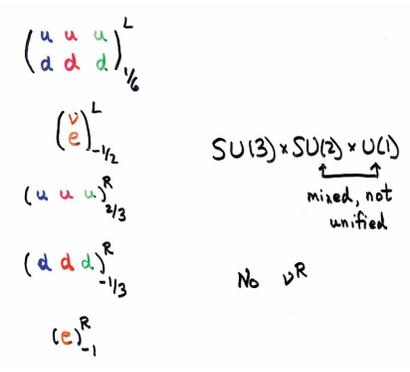}
\caption{Gauge structure of matter}
\label{fig3}
\end{center}
\end{figure}

By demanding full quantum-mechanical consistency, specifically the cancellation
of gauge symmetry anomalies, we can derive constraints among the
hypercharges that
very much improve the situation.  Even at this level, however, there
is no theoretical
barrier to prevent a small admixture proportional to B-L into the
electric charge
operator, for example. In the minimal standard model there is a cubic
anomaly in B-L
itself, but that can be cancelled by including a standard model
singlet, which we can
identify as a right-handed neutrino.  (It is interesting to note that
even a tiny electric
charge for the neutrino would forbid neutrino-antineutrino
oscillations, or in other
words Majorana mass terms, and conversely.  This circumstance adds fundamental
interest to an unmet experimental challenge, to determine whether the neutrino
masses are of Majorana type.)

\subsection{Unification}

These two shortcomings of the gauge system in standard model, that is
the occurrence
scattered of multiplets and peculiar hypercharges, are both overcome
quite beautifully
if we are willing to postulate larger gauge symmetry (spontaneously broken, of
course). With the natural embedding of $SU(3)\times SU(2)\times U(1)$ into
$SU(5)$ we find the
fermions of a single family fall into just two multiplets, a
conjugate vector $\bar {\bf 5}$
plus an  antisymmetric tensor $\bf {10}$.  Moreover their
hypercharges are uniquely
determined, and in agreement with experiment. The ugly ducklings of
the standard
model, upon unification, mature into graceful swans. The ultimate in
unification, while
not addressing family replication, is obtained with the slightly
larger gauge symmetry
$SO(10)$, as shown in Figure 4.  Now the fermions are all in a single
multiplet, the spinor
$\bf{16}$.   The additional particle needed to fill out this mutiplet
is a right-handed
neutrino.  It plays an important constructive role in the theory of
neutrino masses, by
allowing the seesaw mechanism.  A zealot might make a case that it
has thereby already
been observed, albeit indirectly.  Be that as it may, the
forced incorporation of a
right-handed neutrino should probably be viewed as an asset rather than an
embarrassment.
\begin{figure}
\begin{center}
\includegraphics[scale=1.5]{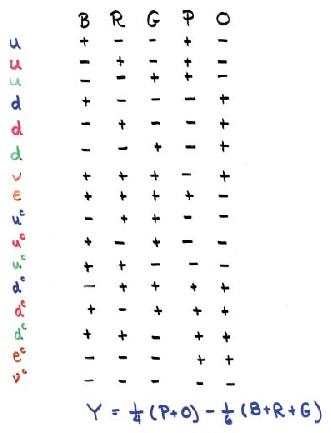}
\caption{Unification of quantum numbers in SO(10)}
\label{fig4}
\end{center}
\end{figure}

At first sight there appears to be a grave difficulty with these
unification schemes, in
that they appear to predict too much, specifically equality of the
strong, weak, and
properly normalized hypercharge couplings, which is definitely not what we
observe.  By now it is becoming a familiar story, depicted in Figure 5, that after
calculating the dynamics
of these theories properly and quantitatively, taking into account
the effects of vacuum
polarization, we find that this apparent difficulty might be resolved
triumphantly.
Indeed, extending the logic of QCD running, already displayed in
Figure 1, to include the
other interactions, we realize that the relative values of the couplings are
scale-dependent.  We should have equality of the couplings only at very short
distances, or high energies, before the asymmetric screening and
anti-screening
clouds distort it.

Following out this idea, we find that by accounting for the renormalization
effects of virtual particles in the minimal standard model (as
usually understood) we get
qualitative but not quantitative agreement.  But if we include expand
the calculation to
implement the effects supersymmetry, in a minimal realization,
beginning at a mass
scale of order 1 Tev, we find quite remarkable agreement, as shown in Figure 5.
Low-energy supersymmetry is an attractive hypothesis for several
other reasons, as
I'll review shortly, but this result, because it is quantitative,
seems to me by far the
most compelling.

The unification of gauge couplings occurs at a very large mass scale, of order
$M_{\rm U} \approx 10^{16}$ GeV, indicating that this is the scale of
unified symmetry
breaking.  It is extraordinary that measurements at accessible
energies, $10^2$ GeV or
less, can point us so specifically to this enormously larger scale.
It happens because the
running of the (inverse) couplings is logarithmic, so that it takes a
lot of leverage to
overcome a modest difference.  The logarithmic running of formally
dimensionless
couplings is a profound consequence of relativistic quantum field
theory in four
space-time dimensions.  The apparent success of this unification
calculation therefore,
on the face of it, suggests that the principles of quantum field
theory continue to be
valid up to energies, or down to distances, many orders of magnitude
beyond where
they were discovered or have been tested directly.
\begin{figure}[t]
\begin{center}
\includegraphics[scale=3]{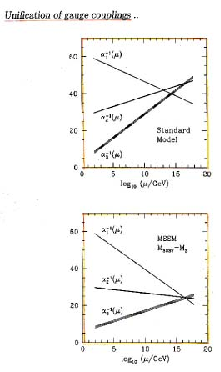}
\caption{Unification of couplings}
\label{fig5}
\end{center}
\end{figure}

The occurrence of a large mass scale $M_{\rm U}$ has important
conceptual advantages.
Unification of gauge interactions inevitably involves putting quarks
and leptons on the
same footing, and upon doing this it becomes difficult to avoid the
occurrence of
significant transitions between them.  From that arises baryon number
violation, and
thus proton decay, unless some conspiracy intervenes.  Baryon number violation
certainly occurs through gauge boson exchange in $SU(5)$ or $SO(10)$,
or any large
gauge symmetry group.  For bringing the predicted rates down to an
acceptable level,
the heavy propagator suppression $M_U^{-4}$ is most welcome.  On the
positive side,
the see-saw mechanism for neutrino masses gives the estimate $m_{\nu} \sim
m_t^2/M_{\rm U}$ for the heaviest observed neutrino mass, where $m_t$
is the top
quark mass, and that's close to what is observed.

It's also most intriguing that $M_{\rm U}$  is close to, although
significantly smaller
than, the Planck mass $G_N^{-\frac{1}{2}} \approx 10^{18}$ GeV.
Gravity responds
directly to energy-momentum, so its effective strength evolves with
the mass scale (for
virtual exchanges) at which it is defined, even classically.  Another
aspect of this is the
Newton constant is, in our usual $h=c=1$ units, a dimensional
coupling.   It is as we
approach the Planck scale that amplitudes involving graviton exchange,
straightforwardly extrapolated, from being much feebler than the
other interactions,
approach quantitative equality.  Independent of any detailed theoretical
implementation, this numerical circumstance strongly supports the
idea that a further
stage of unification, in which both gauge interactions and gravity
take part, is a physical
reality.

\subsection{Low Energy Supersymmetry}
Low-energy supersymmetry has several other important advantages, besides its
helpful role in the quantitative aspect of unification.

Low-energy supersymmetry protects the Higgs (mass)$^2$ term, which governs
the scale of electroweak symmetry breaking, from quadratically
divergent radiative
corrections.  As long as the scale of mass splittings between
standard model particles
and their superpartners is less than a Tev or so, the radiative
corrections to this
(mass)$^2$ are both finite and reasonably small.  (In detail, things
are not quite so clean
and straightforward; there is the ``$\mu$ problem'', which is a very
interesting and
important subject, but too intricate to discuss here.)

This general qualitative relationship between mass splittings of supersymmetry
multiplets and the observed weak scale penetrates also has a more specific and
quantitative aspect.  Supersymmetry relates the physical mass of the lightest,
``standard model-like" Higgs particle, which in the absence of
supersymmetry is a free
parameter, to the masses of W and Z bosons.  There is some model
dependence in this
relationship. But within minimal or reasonably economical
supersymmetric extensions
of the standard electroweak model the Higgs mass is generally
predicted to be near -- or
below! -- existing experimental limits, as shown in Figure 6.  This
renders the models
subject to quick falsification at LHC or, more optimistically, to
fruitful vindication. With
vindication would come the emergence of a rich Higgs-sector
phenomenology starting
somewhere below 150 GeV.

\begin{figure}[t]
\begin{center}
\includegraphics[scale=4]{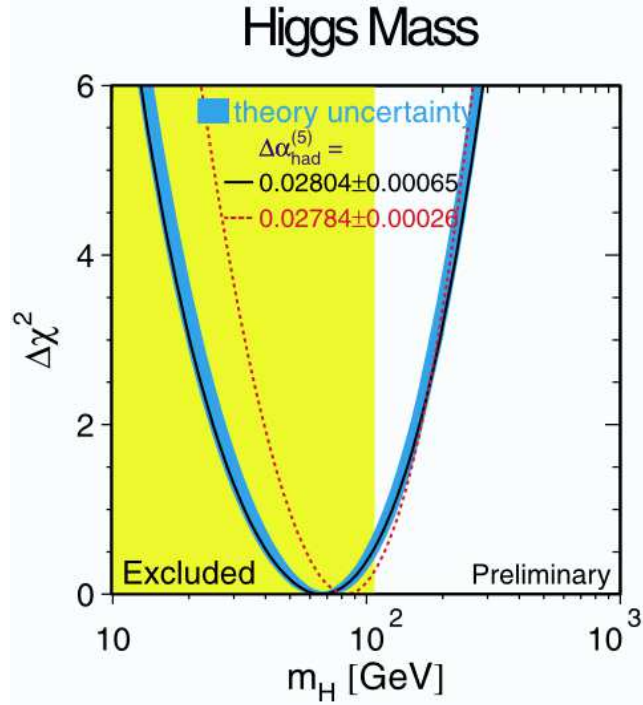}
\caption{Expectations for the Higgs particle mass}
\label{fig6}
\end{center}
\end{figure}

The optimistic scenario gains credibility from another advantage of
supersymmetry.
This is the important though negative virtue, that for precision electroweak
measurements supersymmetric extensions of the standard model
generally yield only
small deviations from the predictions of the standard model itself.
That's a good thing,
because the standard model agrees remarkably well with these measurements.  The
situation is depicted in Figure 7.   Several large classes of rival models to
low-energy
supersymmetry associate electroweak symmetry breaking with new strong
interactions.  In these models, which include Technicolor and both in
its original form
and in its extra-dimensional disguises, radiative corrections to the
Higgs (mass)$^2$ are
rendered finite by form-factors, rather than cancellations; and
though the additional
radiative contributions in these models are finite, there is no
general reason to expect
that they are especially small.  Indeed, to the extent that they
support specific
calculations, one finds that such models generically have severe difficulty in
accommodating existing precision measurements.
\begin{figure}[t]
\begin{center}
\includegraphics{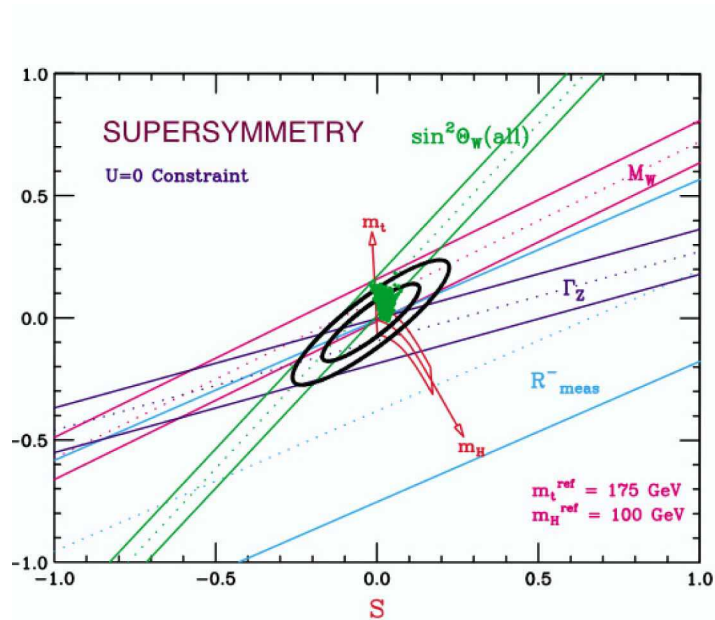}
\caption{Comparison of supersymmetry models to precision electroweak data}
\label{fig7}
\end{center}
\end{figure}

Finally, low-energy supersymmetry can provide an excellent candidate to provide
the dark matter of cosmology.  It's plausible that the lightest
particle with odd
$R$-parity, where $R \equiv  (-)^{B+L+J}$ is stable on cosmological
time scales, because
the quantum numbers that go into the definition of $R$ are well
respected.   The lightest
$R$-odd particle, usually called the LSP (Lightest Supersymmetric
Particle) could be
some linear combination of the photino, zino, and Higgsino.  Indeed,
the production of
these particles in big bang cosmology is about right to account for
the observed density
of dark matter.

\section{Gravity}
General relativity manifestly provides a beautiful, conceptually
driven theory of
gravity.   It has scored many triumphs, both qualitative (big bang
cosmology, black hole
physics) and quantitative (precession of Mercury, binary pulsar).
The low-energy
effective theory of gravity together with the other interactions is defined
algorithmically by the minimal coupling prescription, or equivalently
by restricting to
low-dimension operators.  Since, in this context, ``low" means
compared to the Planck
energy scale, this effective theory is very effective indeed.  We can
and do obtain
unambiguous, apparently accurate answers in the applications above
and many others
by using this theory.  And it is perfectly quantum-mechanical,
supporting for example
the existence of gravitons as the particulate form of gravity waves.

What makes this very tight, predictive, and elegant theory of quantum
gravity ``bad"
is not that there is any experiment that contradicts it.  There
isn't.  Nor, I think, is the
main problem that this theory cannot supply predictions for totally
academic thought
experiments about ultrahigh energy behavior.  It can't, but there are
more pressing
issues, that might have more promise of leading to contact between theory and
empirical reality.

A great lesson of the standard model is that what we have been
evolved to perceive as
empty space is in fact a richly structured medium. It contains
symmetry-breaking
condensates associated with electroweak superconductivity and
spontaneous chiral
symmetry breaking in QCD, an effervescence of virtual particles, and
probably much
more.  Since gravity is sensitive to all forms of energy it really
ought to see this stuff,
even if we don't.  A straightforward estimation suggests that empty
space should weigh
several orders of magnitude of orders of magnitude (no misprint
here!) more than it
does.  It ``should" be much denser than a neutron star, for example.
The expected
energy of empty space acts like dark energy, with negative pressure,
but there's much
too much of it.

To me this discrepancy is the most mysterious fact in all of physical
science, the fact
with the greatest potential to rock the foundations.  We're obviously
missing some
major insight here.  Given this situation, it's hard to know what to
make of the
ridiculously {\it small\/} amount of dark energy that presently
dominates the Universe!

Another disappointing aspect of our effective theory of gravity is
that it walls off the
description of gravity from the theory of the other interactions.
The minimal coupling
procedure for incorporating gravity can accommodate any quantum field
theory of the
rest of Nature; it neither constrains nor significantly modifies the
physical content of
such theories.  Thus it fails to live up to the promise of the
unification of couplings
calculation, with its pointer to the Planck scale, as I discussed above.

 From these perspectives, a profoundly exciting aspect of
supersymmetry is that the
extension of non-gravitational (flat-space, rigid) supersymmetry to
supergravity is
considerably more complicated and delicate than straight minimal
coupling.  One must
include very specific additional terms, some of which do not include
gravitons or
gravitinos. Indeed, these terms play important roles in attempts to construct
phenomenologically viable models realizing low-energy supersymmetry. More
specifically, the non-minimal terms are used, in conjunction with a
gravitational
``hidden sector" to generate soft supersymmetry-breaking mass terms, or a small
$\mu$ term.   People have discussed hidden sectors that derive from
structures living
elsewhere in extra spatial dimensions, or from additional gauge
structures as suggested
by the heterotic string.  It's remarakable that there are genuine
prospects for accessing
the deep structure of gravity, and maybe even discerning evidence for
these other
exotic structures, by experiments at accelerators!

\section{The Flavor/Higgs Sector}
The third sector consists, one might say, of the potential energy
terms.  They are the
terms that don't arise from gauge or space-time covariant
derivatives.  (Note that field
strengths and curvatures are commutators of covariant derivatives.)
All these terms
involve the Higgs field, in one way or another.  They include the
Higgs field mass and its
self-coupling, and the Yukawa couplings.   We know of no deep
principle, comparable to
gauge symmetry or general covariance, which constrains the values of
these couplings
tightly.

For that reason, it is in this sector where continuous parameters
proliferate, into the
dozens.  Basically, we introduce each observed mass and weak mixing angle as an
independent input, which must be determined empirically.  The
phenomenology is not
entirely out of control: the general framework (local relativistic
quantum field theory,
gauge symmetry, and renormalizability) has significant consequences,
and even this
part of the standard model makes many non-trivial predictions and is highly
over-constrained.  In particular, the Cabibbo-Kobayashi-Maskawa (CKM)
parameterization of weak currents and CP violation has, so far,
survived close new
scrutiny at the B-factories intact.

Neutrino masses and mixings can be accommodated along similar lines,
if we expand the
framework slightly.  The simplest possibility is to allow for minimally
non-renormalizable (mass dimension 5) ``ultra-Yukawa" terms.  These
terms involve
two powers of the scalar Higgs field.  To accommodate the observed
neutrino masses and
mixings, they must occur with very small coefficients.

The flavor/Higgs sector of fundamental physics is its least satisfactory part.
Whether measured by the large number of independent parameters or by the small
number of powerful ideas it contains, our theoretical description of
this sector does not
attain the same level as we've reached in the other sectors.  This
part really does
deserve to be called a ``model'' rather than a ``theory''.

There are many opportunities for experiments to supply additional information.
These include determining masses, weak mixing angles and phases for quarks more
accurately; the same for neutrinos; searches for $\mu \rightarrow
e\gamma$ and allied
processes; looking for electric dipole moments; and others.  The big
question for theorists
is: What are we going to do with this data?

\subsection{Breakout Possibilities}
It is very important to gather all that information, and in the process
of doing so we might
very well find direct evidence for new physics ``beyond the standard
model", which
would be great fun. But I don't find it plausible that pinning down
quark masses and
the CKM matrix, or their leptonic analogues, or even discovering
evidence for new
sources of flavor change and CP violation, will give us the sort of
impetus we'll need to
break through to a theory that's better than, as opposed to just
larger than, the third
(Ugly) system as we know it.

If low-energy supersymmetry is indeed discovered, there will be many additional
masses, mixings, and phases to sort out. There are profound questions
to be answered
here. The dark side of low-energy supersymmetry is that it offers
many new potential
sources of flavor and CP violation, including baryon number
violation, which Nature has
made surprisingly sparing use of.  Some of these are associated with
low-dimension
operators, so that {\it a priori\/} they are sensitive to physics at
high mass scales,
where exotic effects of quantum gravity and exchange of the new
particles associated
with unification are unsuppressed.

So various special mechanisms and symmetries have been postulated, to keep
the basic
ideas of low-energy supersymmetry and unification phenomenologically
viable. None
is uniquely convincing, and we will surely need experimental guidance
to figure out
which, if any, is on the right track.

Here's a specific example of an exciting and bizarre, but not
altogether gratuitous or
  crazy possibility.  There's a parameter $m_0$, the universal soft
mass term, that
appears in models of low-energy supersymmetry. It's a kind of fudge factor, and
several theoretical ideas suggest that putting $m_0 =0$ might be
desirable.  But
concrete implementations of this idea appear to run into severe
phenomenological
problems.  Specifically, they predict that the LSP is a charged
particle, the stau; but
cosmology puts severe constraints on the mass density in any new stable charged
particles.

Now there's a very interesting possibility to evade this problem. It
could be that what
appears to be the LSP, if we take into account only the
supersymmetric partners of
standard model particles, is not truly the lightest $R$-parity odd
particle in Nature.
Gravitinos, axinos, or some other weakly coupled ``inos" might be
lighter.  The staus
would decay into them.  In that case, the stau lifetime could be very
long by particle
physics standards, but short on cosmological scales.  These charged
particles could leave
long tracks, and one might even have difficulty observing that they
decay at all!

Finally let me mention one redeeming virtue of the Higgs sector.
(``Virtue" might be too strong; actually, what I'm about to do is
more in the nature of
advertising a bug as a feature.)   The Lagrangian of the standard model, due to
constraints of gauge symmetry, is constructed almost entirely from
hard (mass dimension 4), strictly
renormalizable terms.  The lone exception is the term responsible for 
the Higgs field (mass)$^2$ term, which is proportional to the operator
$\phi^\dagger \phi$.

Now let's entertain the notion that there might be reasonably light,
standard-model
singlet particles deriving from physics at high energy scales.  How
could they couple to
the particles we know?   Since the constraints of gauge symmetry
still operate, the
couplings of these new particles have to be built on top of the
couplings that appear in
the standard model.  Therefore -- with one exception -- they will
have mass dimension
greater than four, and they will be suppressed by coefficients in
which inverse powers
of the high scale appears.  The exception, of course, is when they
couple to Higgs
particles.   So the Higgs field is uniquely susceptible to this sort
of exotic coupling.

Light particles of the sort just mentioned could arise as the
Nambu-Goldstone bosons
of spontaneous symmetry breaking, but in that case they'd be
derivatively coupled,
and the derivatives boost the mass dimension of possible couplings
back up past four.
More interesting from this perspective are moduli fields
associated with flat
directions in supersymmetry models.  In the limit of exact
supersymmetry they are
massless but {\it not\/} derivatively coupled; if the scale of
supersymmetry breaking is
low, they could remain light enough to be accessible -- but only
through the Higgs
sector!

\section{Outlook/Apology}
In keeping with my assignment here, I've focused on fundamental
issues that have
some fairly direct connection with the LHC program. Although this
brief discussion has
necessarily been quite selective, even given that assignment, I trust
that it has served
to emphasize that this program has extraordinary promise to advance our
understanding of Nature in truly fundamental ways.  Of course, there are other,
complementary ways we can hope to gain fundamental insights. I've
discussed some of
them recently in a similar style elsewhere \cite{spacePart}.  But the
LHC program is
uniquely powerful and sure-fire.

\bigskip

{\small The ideas about unstable LSPs and moduli fields allude to
ongoing work with J.
Feng and with B. Patt and D. Smith, respectively, which will appear
shortly.  I thank
them for discussions and inspiration. }

\end{document}